\documentstyle[11pt,newpasp,twoside,epsf]{article}
\def\edcomment#1{\iffalse\marginpar{\raggedright\sl#1\/}\else\relax\fi}
\reversemarginpar
\def\ls{{_<\atop^{\sim}}}
\def\gs{{_>\atop^{\sim}}}
\def\cgs{ ${\rm erg~cm}^{-2}~{\rm s}^{-1}$ } 

\begin{document}
\title{Optical identification of sources from the HELLAS2XMM 
survey}
\author{Fabrizio Fiore}
\affil{Osservatorio Astronomico di Roma, via Frascati 33, Monteporzio,
I00040, Italy}
\author{Giorgio Matt, Fabio La Franca, G. Cesare Perola}
\affil{Universit\'a Roma Tre, Roma, Italy}
\author{Marcella Brusa, Andrea Comastri, Marco Mignoli, Paolo Ciliegi}
\affil{Osservatorio Astronomico di Bologna, Italy}
\author{Paola Severgnini, Roberto Maiolino}
\affil{Osservatorio Astrofisico di Arcetri, Italy}
\author{Alessandro Baldi, Silvano Molendi}
\affil{IFC/CNR, Milano, Italy}
\author{Cristian Vignali}
\affil{Dept. of Astronomy and Astrophysics, Penn State University, USA}

\begin{abstract}
We present preliminary results on the optical identifications of
sources from the High Energy Large Area Survey performed with
XMM-Newton (HELLAS2XMM).  This survey covers about 3 square degrees of
sky down to a 2-10 keV flux of $7\times10^{-15}$ \cgs (Baldi \&
Molendi these proceedings, Baldi et al. 2001).  The survey good
sensitivity over a large area allows us to confirm and extend previous
Chandra results about a population of X-ray luminous but optically
dull galaxies emerging at 2-10 keV fluxes of $\approx 10^{-14}$ \cgs.
Although the statistics are still rather poor, it appears that at these
X-ray fluxes the fraction of these galaxies is similar or even higher than
that of narrow line AGN.
\end{abstract}

\section{Introduction}

Thanks to their revolutionary capabilities (i.e. arcsec quality
imaging, implying a position reconstruction better than a few arcsec,
and large throughput), Chandra and XMM-Newton have opened up a new
volume of discovery space: a factor 50 increase in sensitivity in the
2-10 keV hard X-ray range.  The deep surveys performed so far cover
only a small field of view: a quarter of a square degree for the
Chandra surveys and a similar area for the XMM-Newton Lockman hole
survey.  Our approach is complementary to these deep pencil beam
surveys in that we plan to cover a different portion of the
redshift--luminosity plane.  Our purpose is to study cosmic source
populations at fluxes where a large fraction of the hard X-ray Cosmic
background (HXRB) is resolved ($\approx50\%$ at $F_{2-10}>10^{-14}$
\cgs, see e.g. Comastri et al. 1995, 2001), but where a) the area
covered is as large as possible, to be able to find sizeable samples
of ``rare'' objects; b) the X-ray flux is high enough to provide at
least rough X-ray spectral information; and c) the magnitude of the
optical counterparts is bright enough to allow, at least in the
majority of the cases, relatively high quality optical spectroscopy,
useful to investigate the physics of the sources.

Our goal is to evaluate for the first time the luminosity function of
hard X-ray selected sources in wide luminosity and redshift ranges.
By integrating this luminosity function we will compute the hard
X-ray luminosity density per unit volume due to accretion as a
function of the redshift. This will then be compared with the history
of the UV luminosity density (proportional to the history of the
star-formation) and to the prediction of models for the evolution of
structures in the universe (following the scheme proposed by
e.g. Fontana et al.  1999).  This may give us a clue on the
correlations between formation and evolution of AGN and supermassive
black holes and formation and evolution of galaxies.

We report here on the preliminary results of the optical
identification of hard (2-10 keV) X-ray sources from a 3 square
degree survey performed using XMM-Newton public fields (HELLAS2XMM
Baldi \& Molendi these proceedings, Baldi et al. 2001). We compare
these results with those obtained by our collaboration using Chandra
fields and with the result of Chandra pencil beam deep surveys (HDFN,
Hornschemeier et al. 2001, SSA13, Barger et al. 2001 and CDFS,
Giacconi et al. 2001, Tozzi et al. 2001, Norman et al. 2001).

\section {X-ray data and optical identifications}

During the last year we have obtained optical spectroscopic
identification of hard (2-10 keV) X-ray selected sources discovered in
5 Chandra fields (Fiore et al. 2000, Cappi et al. 2001). We are now
complementing our medium deep survey with 15 XMM-Newton fields (Baldi
et al. 2001), for a total of 3 deg$^2$ of sky surveyed at a 2-10 keV
flux limit of about $10^{-14}$ \cgs. 495 sources have been detected in
the 2-10 keV band (at a threshold probability of $2\times10^{-5}$, see
Baldi et al. (2001).

Four of the 15 XMM-Newton fields (in addition to the Lockman hole,
Hasinger et al. 2001) have been followed-up in the optical band so far
using the ESO 3.6m and the TNG telescopes (the PKS0312-77,
PKS0537-28, Abell2690 and G158-100 fields). In the following we concentrate
on the results obtained for the 115 sources detected in these four
fields. Their 2-10 keV fluxes range between $7.7\times10^{-15}$ and
$10^{-13}$ \cgs.  24 sources (20\%) have $F_{2-10 keV}
\gs 5\times 10^{-14}$ \cgs. Sources with these fluxes should provide
spectra with $\gs$ a few hundred counts in relatively short
($\sim10$ks) XMM-Newton observations. Accurate information on their
spectral shape is therefore either already available or it can be
obtained in the near future. We note that $5\times 10^{-14}$ \cgs
is also the flux limit of the BeppoSAX HELLAS survey, and therefore
``all'' HELLAS sources are well suited for X-ray
follow-up. Conversely, no source from the deep, pencil beam surveys
performed by Chandra has a 2-10 keV flux higher that this limit. The
X-ray follow-up of the faint Chandra sources must await high
throughput missions like Constellation X and XEUS.

\begin{quote}
\begin{figure}
\plottwo{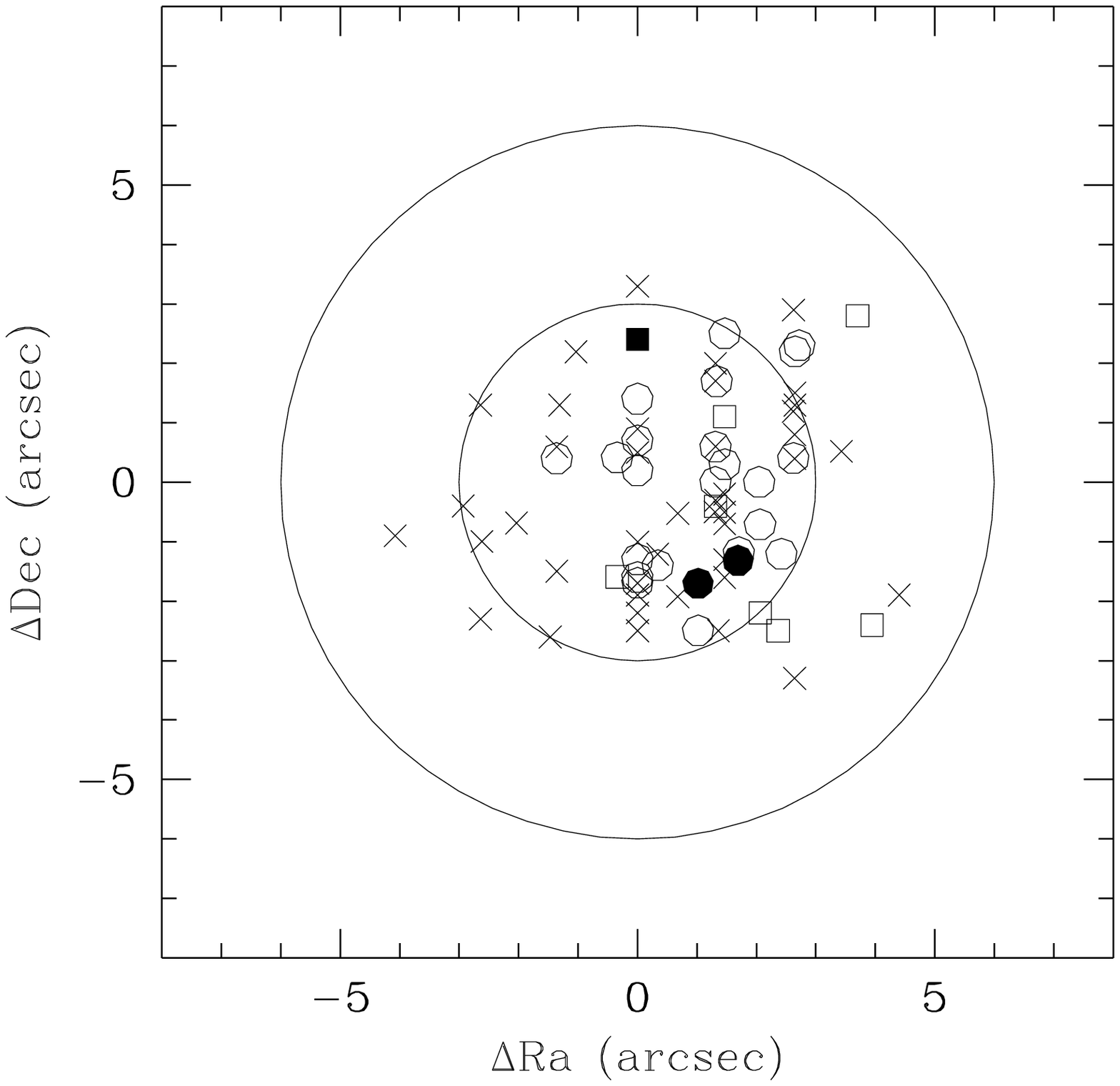}{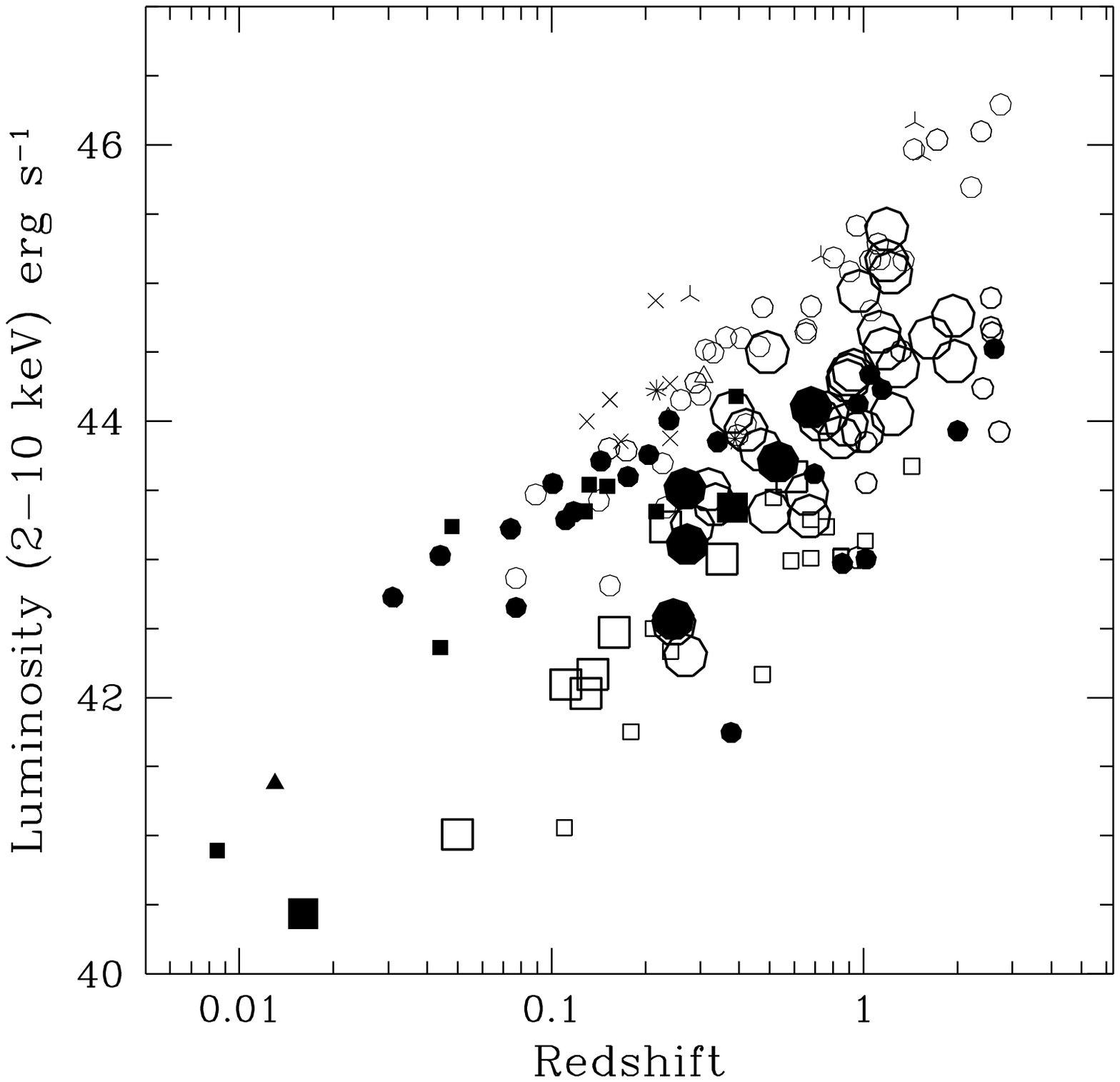}
\caption{left: The deviation in RA and dec between the XMM-Newton positions and
the position of the nearest optical source. 
Figure 2. right: The luminosity-redshift diagram for the HELLAS2XMM sources 
(big symbols); the BeppoSAX HELLAS sources (smaller symbols at lower
redshifts); and the Chandra SSA13 and HDFN deep surveys (smaller symbols
at higher redshifts, data from Barger et al. 2001).
In both diagrams different symbols identify different source classes:
Open circles = broad line, quasars and Sy1; 
filled circles = narrow line AGN; filled squares = 
starburst galaxies and LINERS; 
open squares = optically `normal' galaxies; stars = clusters of
galaxies; open triangles = radio loud quasars; skeleton triangles = 
BL Lacertae objects.
In the left panel crosses mark unidentified sources.
}
\end{figure}
\end{quote}

We have deep R band images for 86 (75\%) of the 115 HELLAS2XMM X-ray
sources. We found optical counterparts brighter than R=24.5 within 5''
from the X-ray position in 75 cases (actually within 3'' in
most of these cases, see figure 1). 11 sources (about 13\%) 
have R$\gs24.5$.  For one fourth of the sources in the sample (34
sources) we have already obtained spectra in an ESO 3.6m run performed
on Dec. 2000.  This sample can be combined with previous works based
on Chandra 2-10 keV detections (see Fiore et al. 2001 and references
therein).  Five of the seventeen Chandra sources with redshifts are
common to the XMM sample.  Therefore the total number of hard X-ray
selected sources with optical spectroscopy is at the moment 46.  This
is one of the largest sample of faint hard X-ray selected sources with
redshifts (see figure 2, which shows the $L_{2-10 keV}$--redshift
diagram for these sources and other hard X-ray selected samples).  We
expect to increase the optical identifications in two 3.6m and 1 TNG
runs already scheduled for the second half of 2001.

\subsection{Source breakdown}

The source breakdown is intriguing: while about half of the sources
are normal broad line quasars, the other half is quite varied: there
are X-ray obscured, emission line AGN, starburst galaxies and, most
unusually, X-ray luminous but apparently ``normal'' galaxy at
$0.05<z<0.35$ (8 out of 46). Their X-ray luminosity, in the range
$10^{42}-2\times10^{43}$ erg s$^{-1}$, is 10-100 times higher than
that predicted based on their optical luminosity, and similar to that
of Seyfert 1 galaxies.  Their X-ray count ratios indicate a hard
X-ray spectrum in many of them.  All this strongly
suggests an (obscured) AGN in these objects (see Comastri et
al. these proceedings and Fiore et al. 2001 for a more detailed
discussion on one of these objects). Possibly obscured AGN emission
lines could be overshined either by the stellar continuum or by a
nuclear non thermal continuum.  Alternatively, they could be not
efficiently produced.  Interestingly, the fraction of narrow line AGN
(5 out 46) is smaller than that of X-ray loud, optically dull
galaxies. A similar result has been recently published by Barger et
al. (2001), based on deep Chandra surveys.  Both increasing the
sample of these objects and sensitive multifrequency follow-ups are
clearly crucial to discriminate among competing possibilities.
The AGN populating hard X-ray surveys may span ranges of X-ray and optical
properties wider than previously thought, with important consequences
for the evaluation of the total `accretion luminosity' of the Universe 
(see e.g. Iwasawa \& Fabian 1999).

\subsection{X-ray to optical ratios}

We have found in our photometric observations that 65\% of the X-ray
sources have a counterpart brighter than R=22, 23\% have
$22<$R$<24.5$ and 12\% have R$>24.5$. It will not be probably
feasible to obtain redshifts for the latter faint sources through
optical spectroscopy. The remaining alternatives are a) to obtain
redshift directly from the X-ray spectra (Baldi et al. 2001b in
preparation); and b) to obtain approximate redshift through
multicolor optical and near infrared photometry.

Figure 3 shows the X-ray to optical flux ratio of the XMM sources as a
function of the X-ray flux. This is compared with the analogous ratios
for the sources from the Chandra SSA13 and HDF-N deep pencil beam
surveys.  Solid lines show loci of constant R magnitude.  We note that
while the fraction of sources with R$\gs24.5$ will be higher in the
deep surveys, their X-flux will be smaller (by a factor of typically
10) and therefore it will be difficult to obtain their redshift
through optical and/or X-ray spectroscopy. This means that a
relatively large fraction of sources from deep surveys will remain
unidentified, at least until the advent of the next generation of
optical and infrared large telescopes (NGST from the space and the 
30m\footnote{see e.g. http://staging.noao.edu/gsmt.html}
and 100m\footnote{http://www.eso.org/projects/owl/} telescopes from
the ground, Gilmozzi \& Dierickx 2000 and references therein).
Conversely, the X-ray fluxes of the XMM 3 degrees survey are high
enough to try to search for iron and oxygen features which in turn
could provide redshifts.

\setcounter{figure}{2}
\begin{quote}
\begin{figure} 
\plottwo{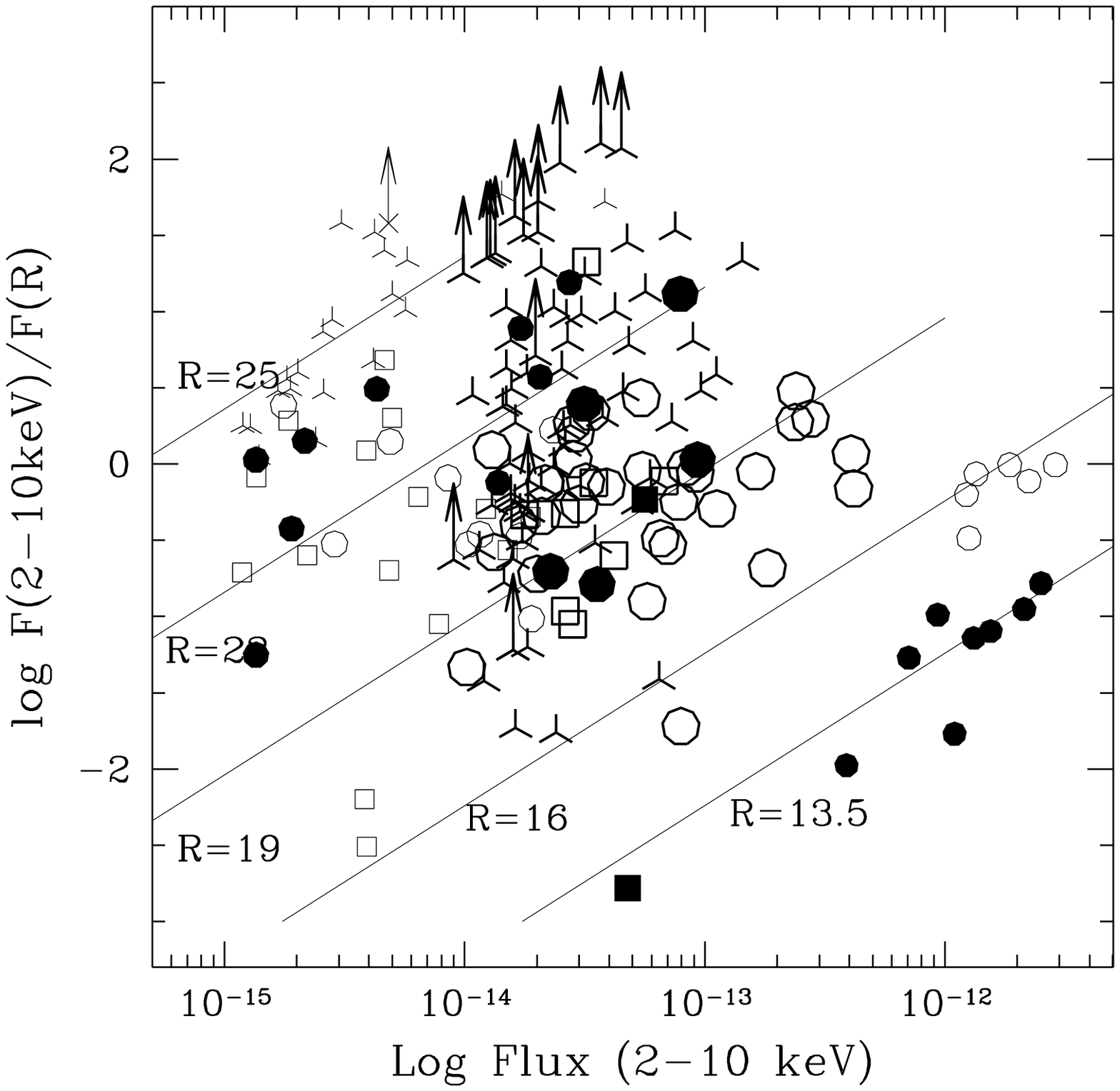}{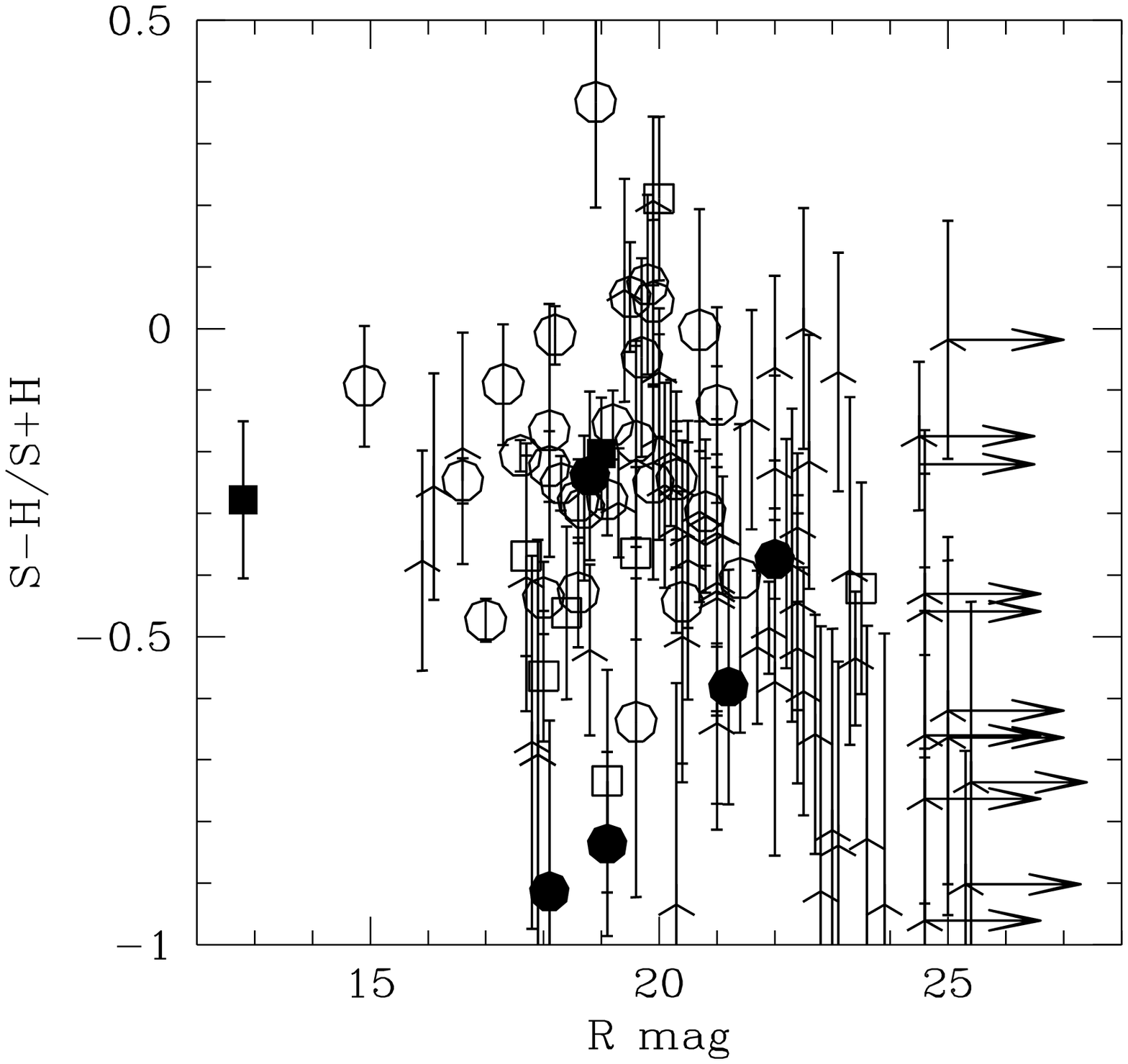}
\caption{left: the ratio between the X-ray (2-10 keV) and optical
(R band) fluxes for the HELLAS2XMM sources (big symbols), the sources
from the SSA13 and HDF-N deep surveys (data from Barger et al.  2001
and Hornschemeier et al. 2001, small symbols at low fluxes) and two
samples of optically selected AGN: PG quasars and Compton thin
Seyfert 2 galaxies (small symbols at high fluxes).  Symbols as in
Figure 1 and 2; skeleton triangles mark unidentified sources.
Figure 4. right: The X-ray softness ratio (S-H)/(S+H) (S=0.5-2keV
flux; H=2-10 keV flux) as a function of the R band magnitude for the
HELLAS2XMM sources}
\end{figure}
\end{quote}

Figure 4 shows the X-ray softness ratio as a function of the R band
magnitude. The figure shows clearly that soft sources tend to have
brighter R magnitudes. The average magnitude of the sources with
((S-H)/(S+H)$>-0.3$) is 19.8 with a dispersion of 2.1, while the
average magnitude of the sources with ((S-H)/(S+H)$<-0.3$) is 21.5
with a dispersion of 2.2. The optical counterparts of most soft
sources is brighter than R$\ls23$, bright enough to measure the
redshift and in turn to evaluate the accretion luminosity. Conversely,
a relatively large fraction of hard sources have R$\gs24.5$. As
mentioned above for many of them it will be difficult to obtain a
redshift and so to estimate the luminosity. It will therefore be
difficult, if not impossible, to localize the redshift from which the
accretion power of these sources is coming and to measure it, using
optical spectroscopy. For the brighter X-ray sources the redshift
could be estimated directly from the X-ray spectrum. This will be
unfeasible for the faint hard sources discovered in deep and ultradeep
surveys.  This means that deep pencil beam surveys are probably less
efficient to depict the history of the accretion in the Universe than
shallower and larger area surveys like the HELLAS2XMM survey.

Figure 3 also illustrate the efficiency of X-ray surveys in probing
accretion (as also often remarked by R. Mushotzky). At a 2-10 keV flux
of, say, $10^{-14}$ \cgs there are 300--400 sources per square degree
(see Baldi et al. these proceedings and references therein).
Optically selected AGN have a 2-10 keV to R band flux ratio of at most
1 (PG quasars, see figure 3), and therefore their R band magnitude at
$F_{2-10 keV}=10^{-14}$ \cgs is R$\ls$21. The number of optically
selected AGN at these magnitudes is 100-150 per square degree, a
factor 2-3 less than hard X-ray selected AGN (also see La Franca et
al. these proceedings).  To reach an AGN surface density of $\sim300$
deg$^2$ optical surveys must be pushed down to B=23.5, where contamination
from compact emission line galaxies is very large, making the optical
spectroscopic follow-up much less efficient than for X-ray surveys
(Mignoli \& Zamorani 2001 in preparation).

\acknowledgments
This research has been partially supported by ASI contracts
ARS--99--75 and I/R/107/00, MURST grants Cofin--99--034 and 
Cofin-00--02--36 and a 1999-2000 CNAA grant.

\end{document}